\documentclass[12pt,preprint]{aastex}

\newcommand{\kms}{km s$^{-1}$~}
\newcommand{\kmss}{km s$^{-1}$}
\newcommand{\vp}{$\upsilon_{p}$~}

\newcommand{\hbeta}{H$\beta$~}
\newcommand{\mgfep}{[MgFe]$^\prime$~}
\newcommand{\hbetaa}{H$\beta$}
\newcommand{\mgfepp}{[MgFe]$^\prime$}
\newcommand{\afe}{[$\alpha$/Fe]~}
\newcommand{\afee}{[$\alpha$/Fe]}

\newcommand{\etal}{et al.~}
\newcommand{\ct}{$(C-T_{1})$~}
\newcommand{\ctt}{$(C-T_{1})$}
\newcommand{\tone}{$T_{1}$~}

\shorttitle{Spectroscopy of the Globular Clusters in M86}
\shortauthors{Park et al.}

\begin{document}
\title{SUBARU Spectroscopy of the Globular Clusters in the Virgo Giant Elliptical Galaxy M86{*}} 

\author{ Hong Soo Park\altaffilmark{1}, Myung Gyoon Lee\altaffilmark{1},
and Ho Seong Hwang\altaffilmark{2} 
}

\email{hspark@astro.snu.ac.kr, mglee@astro.snu.ac.kr,
and hhwang@cfa.harvard.edu  
}

\altaffiltext{1}{Astronomy Program, Department of Physics and Astronomy, Seoul National University, Seoul 151-742, Korea}
\altaffiltext{2}{Smithsonian Astrophysical Observatory, 60 Garden Street,
Cambridge, MA 02138, USA}
\altaffiltext{*}{Based on data collected at Subaru Telescope, which is operated by the National Astronomical Observatory of Japan.}
 
\begin{abstract}

We present the first spectroscopic study of the globular clusters (GCs) 
  in the giant elliptical galaxy (gE) M86 in the Virgo cluster.
Using spectra obtained in the Multi-Object Spectroscopy (MOS) mode of the 
  Faint Object Camera and Spectrograph (FOCAS) on the Subaru telescope,
  we measure the radial velocities for 25 GCs in M86.
The mean velocity of the GCs is derived to be $\overline{v_p}=-354^{+81}_{-79}$ \kmss, which is
  different from the velocity of the M86 nucleus 
  ($v_{\rm gal}=-234\pm 41$ \kmss).
We estimate the velocity dispersion of the GCs, $\sigma_p=292^{+32}_{-32}$ \kmss,
  and find a hint of rotation of the M86 GC system.
A comparison of the observed velocity dispersion profiles of the GCs and stars
  with a prediction based on the stellar mass profile
  strongly suggests the existence of an extended dark matter halo in M86.
We also estimate the metallicities and ages for 16 and 8 GCs, respectively.
The metallicities of M86 GCs are in the range $-2.0<$ [Fe/H] $<-0.2$ 
   with a mean value of $-1.13 \pm 0.47$. 
These GCs show a wide age distribution from 4 to 15 Gyr.

\end{abstract}


\section{Introduction}

Globular clusters (GCs) 
  are excellent tracers
  for studying the formation history of their host galaxies \citep{lee03, bro06}.
In particular, 
  a giant elliptical galaxy (gE) contains thousands of GCs, located close to the galaxy center to very far away in the outer halo.
Therefore, GCs can be used as powerful test particles
  for studying the kinematics and chemical evolution of gE halos.
     
Up to now, there are several kinematic studies of the GC systems in nearby gEs:
M49 \citep{zep00,cot03},
M60 \citep{pie06, bri06,lee08b, hwa08}, 
M87 \citep{coh97,kis98b,cot01}, 
NGC 4636 \citep{sch06, cha08, par10, lee10a}, 
NGC 1399 \citep{kis98, min98, kis99, ric04, ric08, sch10}, 
NGC 5128 \citep{pen04a, pen04b, woo07}, and
NGC 1407 \citep{rom09}. 
The data used for these studies were compiled and reanalyzed in \citet{lee10a},
who 
found that the kinematic properties of the GC systems are diverse among the gEs,  indicating diverse merging and accretion histories of gEs 
  (see also the recent study on the M87 GC system \citep{str11}).

There are also several studies focusing on the spectroscopic ages and metallicities of the gE GCs:
\citet{coh98} for M87,
\citet{bea00} and \citet{coh03} for M49, 
\citet{pen04b}, \citet{bea08}, \citet{woo10}, and \citet{woo10b} for NGC 5128,
\citet{pie06} for M60,
\citet{cen07} for NGC 1407, and
\citet{kis98} and \citet{for01} for NGC 1399.
Recently, \citet{par12a} presented a study of spectroscopic ages and metallicities 
  for the GCs in NGC 4636.
They also compiled the data of all the gE GCs in the literature, and found that
  the GC metallicity distribution in the combined gE sample is bimodal.

We have been carrying out a project to investigate the spectroscopic
  properties of GCs in nearby galaxies
  to understand the formation of the GC systems in galaxies. 
Our study on the kinematics of the GC system of the Virgo gE M60 was
  presented in \citet{lee08b} and \citet{hwa08} and
  that on the GC system of the spiral galaxy M31 
  in \citet{kim07} and \citet{lee08c}.
Recently, we presented measurements of the radial velocities for
  the GCs in NGC 4636 \citep{par10}, 
  and a detailed kinematic analysis of these data in \citet{lee10a}.
We also investigated the chemical properties of NGC 4636 GCs and 
  other gE GCs \citep{par12a}. 
  
%
Here, we present a spectroscopic study of the GCs in M86 (NGC 4406),
  a gE in the Virgo cluster.
This galaxy is one of the best targets for the spectroscopic study of the GC system
  because it is located close to the center of the Virgo cluster
  and harbors GCs out to large radii from the galaxy center \citep{lee10b}.
To date, there has been no published spectroscopic study of M86 GCs.
%

In contrast to the absence of spectroscopic studies,
  there have been several studies of the photometric properties of M86 GCs.
For example,  
  \citet{kun01} and \citet{lar01} found that the color distribution
  of M86 GCs is  bimodal from an analysis of HST/ WFPC2 images. 
This bimodality was confirmed by HST/ACS data \citep{pen06} and 
   ground-based wide-field imaging data \citep{rho04,par12b}.
The radial number density profile of the M86 GC system is approximately well fitted by a de Vaucouleurs law and power law \citep{rho04,par12b}.
Basic information about M86 is summarized in Table \ref{tab-m86gal}.

This paper is composed as follows.
Section 2 describes the spectroscopic target selection, observation, and data reduction.
In \S3, we identify genuine M86 GCs and list the corresponding
photometric and spectroscopic data.
We present the kinematic properties of  the M86 GC system in \S4, and the metallicities and ages of the M86 GCs in \S5.
The primary results are summarized in the final section. 
 

\section{Observation and Data Reduction}\label{data}

\subsection{Spectroscopic Target Selection}\label{datasel}

We selected the spectroscopic targets from a photometric sample of the GC candidates in M86
  identified in 
  Washington $CT_1$ images ($15.8\arcmin\times15.8\arcmin$) 
  taken at the KPNO 4 m telescope \citep{par12b}.
The GCs in M86 appear as point sources in the ground-based KPNO images, and
we first selected point sources around M86 with colors of $0.9<(C-T_1)<2.1$
  as GC candidates. 
This $(C-T_1)$ color selection criterion is 
 effective in selecting
  GC candidates in early-type galaxies: 
  the success rate of the photometric search for GCs 
  is about 90\% in the case of M87 \citep{cot01}, M49 \citep{cot03}, and NGC 4636 \citep{par10}.
  %
We then selected from the bright sources with magnitudes  $19<T_1<21.5$ 
  as the spectroscopic targets.
A total of 67 targets were chosen, which included
  the M86 nucleus and two known faint galaxies (NGC 4406B, VCC 0833). 
We also observed two red, bright point sources with $(C-T_1)\sim 3$ to fill the mask gaps.

\subsection{Observation}

\begin{figure}
\epsscale{1.0}
\plotone{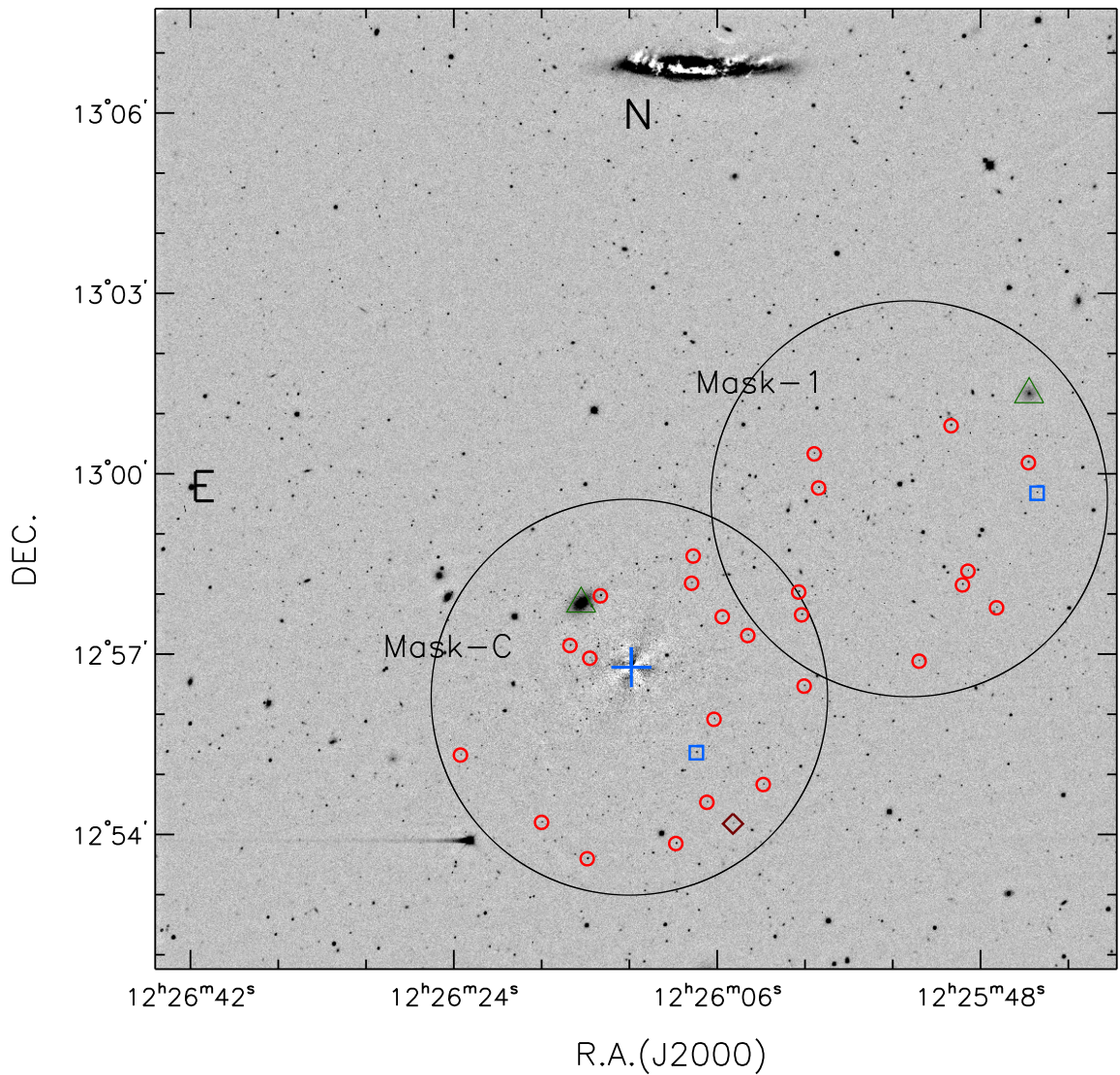}
\caption{Grayscale map of the $T_1$ image of M86 taken with the KPNO 4 m telescope.
The M86 stellar light is subtracted from the original image to highlight the point sources.
The large circles and plus sign indicate the observed masks and the center of M86, respectively.
The small circles, diamond, triangles, and squares represent the GCs in M86,
a probable intracluster GC in the Virgo cluster, background galaxies, and foreground stars, respectively. 
\label{fig-t1image}}
\end{figure}

We obtained spectra of the 67 targets 
  from observations in the Multi-Object Spectroscopy (MOS) mode of 
  Faint Object Camera and Spectrograph (FOCAS; \citealt{kas02}) 
  at the Subaru 8.2 m telescope on 2002 April 21.
We observed two circular masks with diameters of $6\arcmin$.
Figure \ref{fig-t1image} presents a grayscale map of 
  the $T_1$ image of M86 taken at the KPNO 4 m telescope,
  which shows the positions of the spectroscopic targets as well as 
  the observed masks.
We subtracted the M86 stellar light 
from the original image
using 
IRAF/ELLIPSE\footnotemark\footnotetext{IRAF is distributed by the
National Optical Astronomy Observatories, which are operated by the
Association of Universities for Research in Astronomy, Inc., under
contract to the National Science Foundation.} 
  task  to show the point sources clearly.

The observational log is given in Table \ref{tab-mask}. 
We used a medium-dispersion blue grism (300B)
 with a dispersion of 1.34 \AA~ pixel$^{-1}$ and  
 the order-cut filter L600 covering $3700-6000$ \AA.
Seeing during the observation was $0.6\arcsec$. 
To make the masks,
  we obtained $R$-band pre-images with exposure times of 180 s  
  in the FOCAS camera mode under a seeing condition of $\sim1.0\arcsec$ on 2002 March 9. 
Using these images, we constructed masks with Mask Design Pipeline (MDP), 
  a software utility for MOS \citep{sai03}. 
The slit width along the dispersion axis was 0.8\arcsec,
  and the resulting spectral resolution was
  $R \sim 500$.
  
We used three 1200 s exposures in mask-C and 
 one 1800 s exposure in mask-1.
We also obtained the comparison spectra with Th-Ar lamps 
  before and/or after each exposure.
We used the FOCAS long-slit spectroscopy mode  
 to calibrate the flux, radial velocity, and metallicity.
We observed a standard star, BD+33d2642, for the flux calibration
 and five Milky Way (MW) GCs (M5, M13, M92, M107, and NGC 6624) 
 for the velocity and metallicity calibrations. 
We observed the MW GCs 
with stepping scan mode to sample an area larger than that covered by the slit by moving the slit along the dispersion direction.
We observed these calibration targets during the same run 
 and also used their spectra in the study of the NGC 4636 GCs \citep{par10}.
Details of the 
long-slit mode observation are 
  given in Section 2.2 and Table 2 of \citet{par10}.
  
\subsection{Data Reduction}\label{datared}

We first applied basic processing techniques 
  (overscan correction, bias subtraction, and cosmic ray rejection)
  to the CCD images using IRAF tasks: 
  the CCD images were obtained with a pair of 4K $\times$ 2K CCDs.
We used the $FOCASRED/bigimage$ task in IDL \citep{sai03}
  to produce 
  a large single image from a pair of CCD images
  and correct distortions in the optics. 
We then clipped the two-dimensional spectrum of each target
out of the single image, and  
  applied a flat-field correction. 
The spectrum from each two-dimensional image 
  was traced, extracted, and sky-subtracted using the IRAF/APALL task. 
We could not extract the spectra of seven
  faint targets because of a low signal-to-noise (S/N) ratio.
Wavelength calibration was performed 
  using the Th-Ar lamp spectra with 
  $\sim$40 useful emission lines in the $3800-6000$ \AA~ range. 
The typical $rms$ error for this calibration is $\sim 0.8$ \AA. 
The flux of the target spectra was calibrated
  using the flux standard star.
  
\begin{figure}
\epsscale{0.5}
\plotone{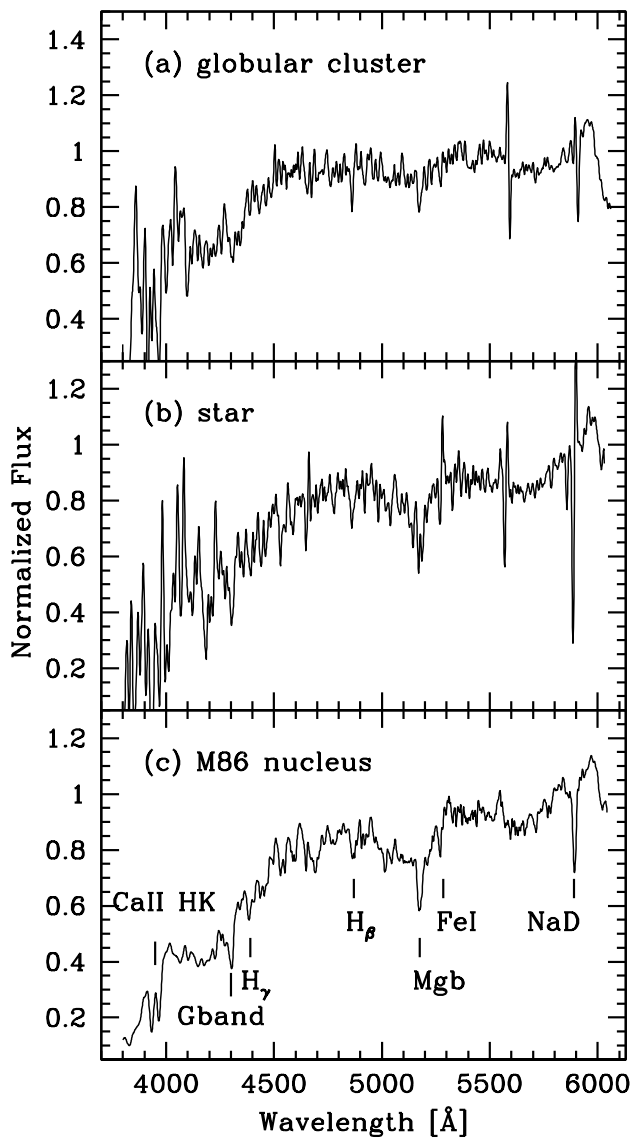}
\caption{ 
Sample spectra : 
         (a) a GC in M86 (ID=284) with $T_1=19.72$ mag, $(C-T_1)=1.49$, and  [Fe/H] $=-0.99$ dex, 
         (b) a star (ID=81) with $T_1=19.89$ mag classified as a K2III red giant star, 
         and (c) the M86 nucleus. 
All spectra are plotted in the rest frame, 
 smoothed using a boxcar filter with 6.7 \AA, 
 and normalized at 5870 \AA.
%
%
\label{fig-spec}}
\end{figure}

Sample flux-calibrated  spectra of
 an M86 GC, a foreground star, and the M86 nucleus 
 are shown in Figure \ref{fig-spec}. 
We classified the target (ID 81) in panel (b) as a K2III star
  because of the broad ($\sim$300\AA) absorption feature around the Mgb index.
This feature is typically seen in the K giant star templates \citep{san02}
  but not in GC templates.
Several absorption lines typically seen in old stellar systems 
  including the G band, H$\beta$, and Mgb are clearly visible in the M86 GC spectrum 
  in panel (a).
Absorption features in the spectrum of the M86 nucleus 
 are much broader than those in the GC and star
 because of its large velocity dispersion.


\section{Membership Determination} 

\subsection{Velocity Measurement}

 We measured the radial velocities for the targets
  using the Fourier cross-correlation task, IRAF/FXCOR \citep{ton79}.
A wavelength range of $4200-5400$ \AA~ was used 
  for the cross-correlation 
  because of the low S/N at $\sim 4000$ \AA~ 
  and the strong night sky emission line [O I] at 5577 \AA. 
During the cross-correlation,
 we fit the continuum of the spectra 
 using a spline function with a $2\sigma$ clipping for the low level 
 and a $4\sigma$ clipping for the high level.
Radial velocities were measured for each target
  using the five MW GC templates;
 velocities for M86 GCs derived from the M5, M13, M92, and NGC 6624 templates 
  are consistent 
  within $1\sigma$, 
  but those from the M107 template differ by $2\sigma$.
Therefore, 
  an error-weighted average of the first four measurements 
  was taken to give the 
  final radial velocity for each target.
The error in the measured radial velocity is estimated as 
  $<\epsilon_v>=(\Sigma\epsilon_i^{-2})^{-1/2}$,
  where $\epsilon_i$ is the error in each measurement.

The final number of targets with measured radial velocities is 31. 
The radial velocities for 36 objects among the original 67
  could not be determined because of the poor quality of the spectra.      
%
We determined the radial velocity of the M86 nucleus to be 
  \vp = $ -234\pm 41$ km s$^{-1}$, which is 
  consistent with the previous measurement \vp = $ -244\pm 5$ km s$^{-1}$ \citep{smi00}.
We also measured the radial velocities for two faint galaxies: 
  \vp = $ 777\pm 38$ km s$^{-1}$ for VCC 0833 and \vp = $ 949\pm 23$ km s$^{-1}$ for NGC 4406B.
  These values are also consistent with previous measurements (Sloan Digital Sky Survey and \citealt{str92}).
%
Errors in our velocity measurements 
  range from 20 to 90 \kms with a mean error of $49\pm16$ km s$^{-1}$.
%

\subsection{Membership Determination}\label{velmem}

\begin{figure}
\epsscale{1.0}
\plotone{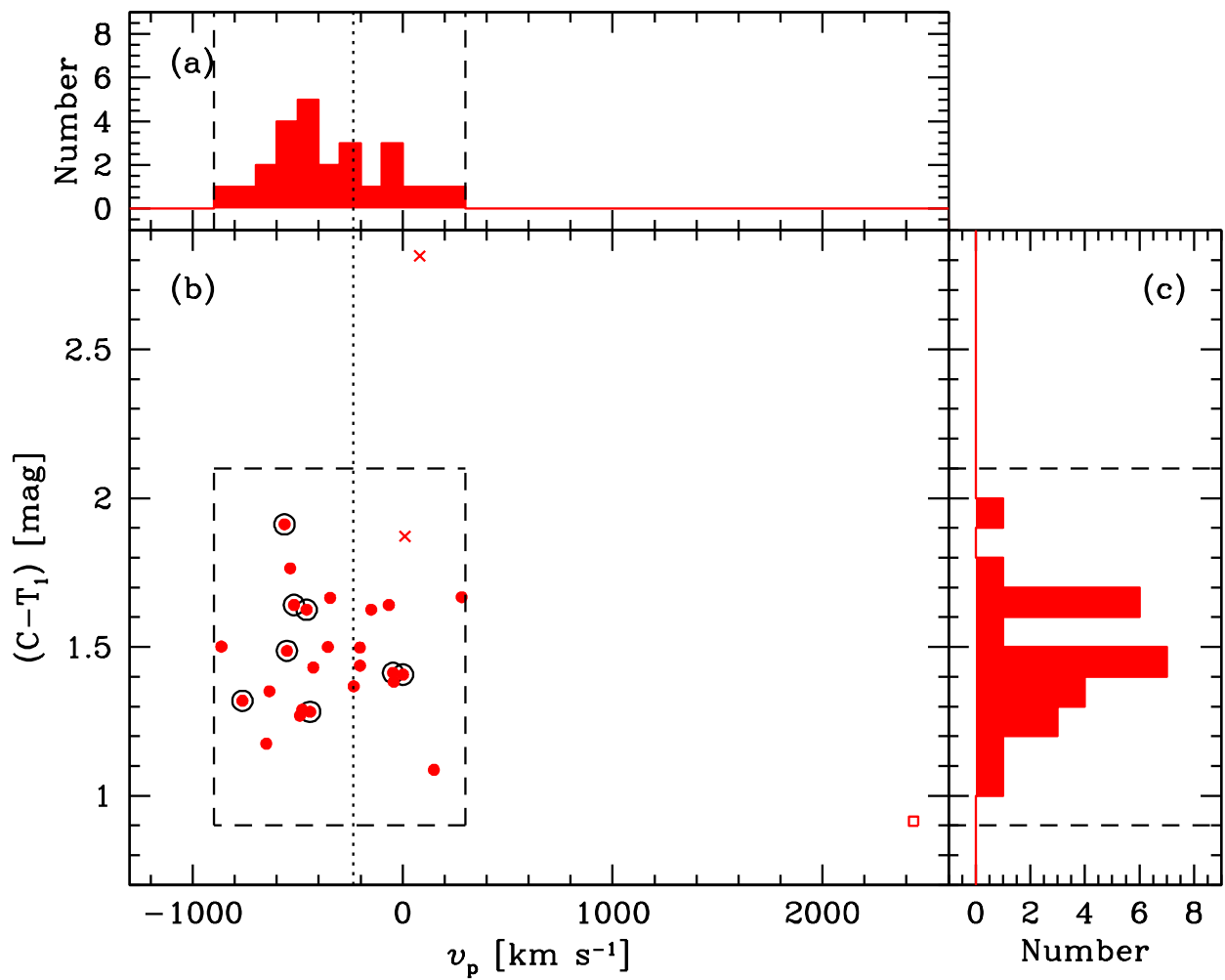}
\caption{ 
\ct color as a function of the radial velocity 
of the M86 GCs :
(a) radial velocity distribution,
(b) \ct versus radial velocity, and
(c) \ct distribution.
The circles, crosses, and square represent the M86 GCs, foreground stars, and
a probable intracluster GC, respectively. 
The open circles indicate the GCs included in the ACSVCS catalog.
The box indicated by the dashed line  represents  the boundary 
for selecting the M86 GCs.
The  vertical dotted line indicates the radial velocity of the M86 nucleus.
%
\label{fig-vpct1}}
\end{figure}

To identify genuine M86 GCs among the 31 objects with measured velocities, 
  we used the \ct colors and the radial velocities.
The \ct colors are plotted as a function of the radial velocity in Figure \ref{fig-vpct1}; 
  note that the M86 nucleus and two faint galaxies are not included in the plot.
The velocity distribution in panel (b) 
  shows that all objects have velocities of --900 to +300 \kms 
  except for one object with very a large velocity of
  \vp $\sim 2400$ \kmss. 
We consider the objects with 
 $-900 <$ \vp $<300$ \kmss~
  and $0.9\le$ \ct $<2.1$
  to be genuine GCs bound to M86 (the velocity of the M86 nucleus is $-234$ \kmss).
The \ctt~ color range is equivalent to a metallicity range 
  of $-$2.24 $\lesssim$~[Fe/H]~$\lesssim$ 0.33 dex
  \citep{lee08a}.

Among the 28 objects (excluding the M86 nucleus and two faint galaxies)
  two do not satisfy the GC selection criteria.
One target (ID 226) is classified as a star
  because its color is too red, \ct $>2.1$,  even though it satisfies the velocity criterion.
The other target (ID 446) satisfies the color criterion 
  but does not satisfy the velocity criterion: \vp $= 2434 \pm 52$ \kmss.
  This object seems to
 be an intracluster GC in the Virgo cluster.

Because the systemic velocity of M86 is similar to the radial velocities
  of the MW stars,
  there could be some MW stars that satisfy our GC selection criteria.
To remove these stars, we performed a careful visual inspection of the target spectra.
We found one object (ID 81) with stellar spectral features (see Section \ref{datared} and Figure \ref{fig-spec} (b))
  and rejected it from the M86 GC catalog.
Thus, 25 genuine M86 GCs out of 28 targets excluding the M86 nucleus and two
faint galaxies were identified.  
We note that there may be one or two more stars among the 25 GCs 
  considering the selection efficiency for the GC candidates (see Section \ref{datasel}). 

The reliability of our M86 GCs was further checked against the ACS Virgo Cluster
Survey (ACSVCS) source catalog \citep{jor09}.
This catalog provides the classification probability for GC candidates in 100 Virgo early-type galaxies.
We found that eight GCs (ID = 448, 284, 65, 270, 316, 107, 430, and 324) among the 25 M86 GCs are included in the ACSVCS catalog, and all of the eight have a GC probability larger than 90\%,
which confirms our classification.   
We show these eight GCs as open circles in Figure \ref{fig-vpct1}.

\subsection{A Catalog of the M86 GCs}\label{speccat}

Table \ref{tab-m86gc} lists the photometric and spectroscopic data 
  including the metallicities of 16 GCs using the BH method 
  and the ages and metallicities of eight GCs using the grid method
  (see Section 5).
The first column represents identification number. 
The second and third columns give the right ascension and declination (J2000), respectively. 
The galactocentric radius and position angle are given in columns 4 and 5, respectively. 
The magnitude and color information in columns 6 and 7 are from \citet{par12b}. 
The eighth column gives the radial velocity and its error measured in this study.
The ninth and tenth columns give the age and metallicity derived from the grid method.
The eleventh column gives the metallicity derived from the BH method.
The final column indicates the corresponding mask from Table \ref{tab-mask}.
The M86 GCs
  are listed first, followed by foreground stars,
  a probable intracluster GC, and the faint galaxies and M86 nucleus.


\section{Kinematics} 

\subsection{Mean Velocities and Velocity Dispersions of the M86 GCs}\label{veldisp}

\begin{figure}
\epsscale{0.5}
\plotone{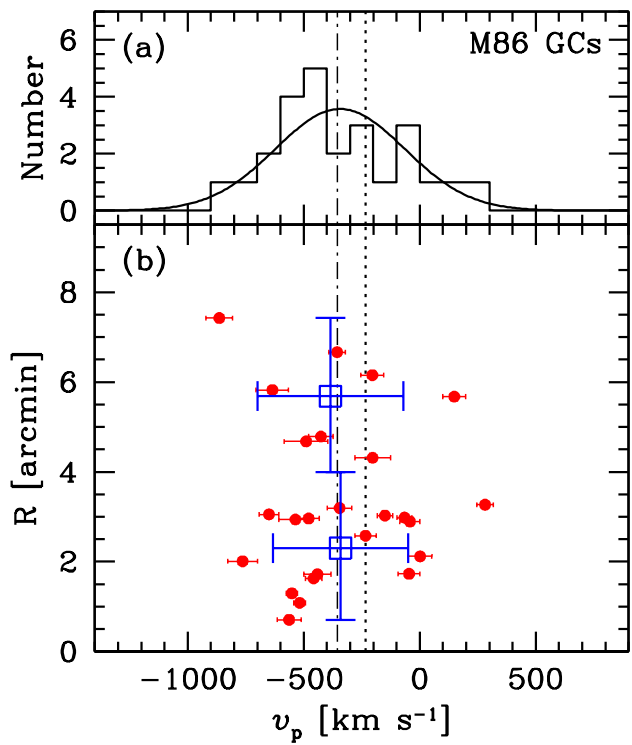}
\caption{(a) Radial velocity distribution 
and (b) projected galactocentric radii versus radial velocities for  the  M86 GCs.
The large open squares represent the mean radial velocities of the GCs in each radial bin 
   represented by the long vertical error bars. 
   The horizontal error bars denote the velocity dispersions in each radial bin. 
The vertical dot-dashed and dotted lines indicate the mean velocity and 
  velocity of the M86 nucleus, respectively. 
The solid curved line in (a) is a Gaussian fit of the data.   
\label{fig-vphist}}
\end{figure}

The M86 GCs in our sample 
  are found in a radial range of 
  42$-$446\arcsec~ (i.e., 3.4$-$36.1 kpc, see Figure \ref{fig-t1image}).
The mean radial velocity for the 25 GCs 
  determined with the biweight location of \citet{bee90}
  is $\overline{v_p}=-354^{+81}_{-79}$ \kmss,
which is smaller than the radial velocity 
  for the M86 nucleus ($v_{\rm gal}=-234\pm 41$ \kmss).
The radial velocity distribution for the M86 GCs is shown in the top panel of Figure \ref{fig-vphist};
  the distribution appears to be Gaussian. 
The $I$ statistics \citep{tea90} gave an $I$ value of 1.022, which is  
  smaller than the critical value for rejecting the Gaussian hypothesis at the 90\% confidence level, $I_{0.90}=1.176$. 
  This suggests that the velocity distribution of the M86 GCs follows a Gaussian distribution.
 A Gaussian fit yields a peak at $v_p=-343$ \kms with a width of $\sigma_p=279$ \kmss. 
%
  
%

The  bottom panel of Figure \ref{fig-vphist} shows the radial velocities 
  as a function of the projected galactocentric radius $R$.
Mean velocities of the GCs 
  in the inner ($42\arcsec \leq R < 240\arcsec$) and 
  outer ($240\arcsec \leq R < 446\arcsec$) regions
  are similar to that of the GCs in the entire region.
The velocity difference between the GC samples and the M86 nucleus
  is more visible in this panel. 

The velocity dispersion of the M86 GCs determined with 
the biweight scale of \citet{bee90} is 
  $\sigma_p=292^{+32}_{-32}$ \kmss. 
The velocity dispersion of the GCs in the inner region, 
  $\sigma_p=292^{+39}_{-39}$ \kmss,
  is similar to that in the outer region, $\sigma_p=314^{+90}_{-90}$ \kmss.
It is noted that 
  the velocity dispersion determined in this study
  could be contaminated by the inclusion of possible MW stars
  because the systemic velocity of M86, $v_{\rm gal}=-234$ \kmss, is
  in the velocity range of the MW stars.

\begin{figure}
\epsscale{1.0}
\plotone{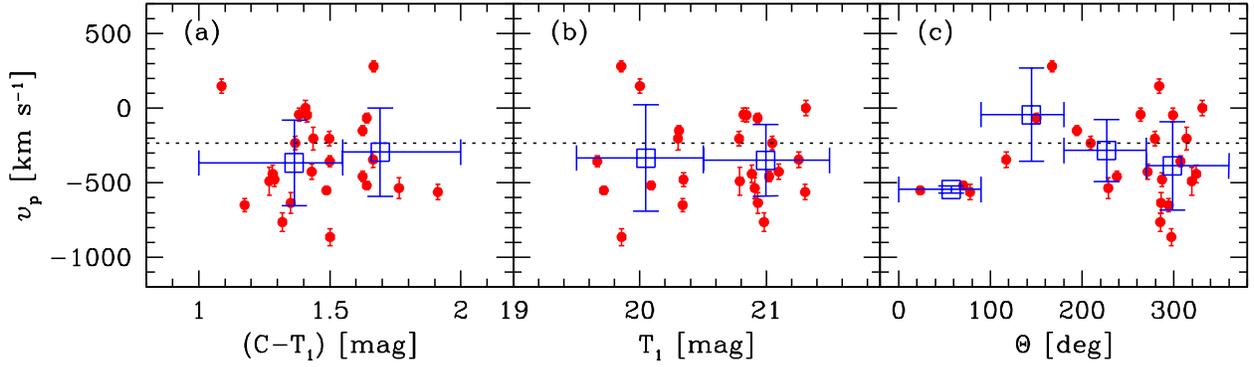}
\caption{ 
Radial velocities as a function of (a) \ct color, (b) \tone magnitude, and (c) position angle, $\Theta$,
for the M86 GCs.
The large open squares indicate the mean radial velocities of the GCs in each bin, 
 represented by a long horizontal error bar. 
The vertical error bars denote the velocity dispersions of the GCs in the radial bins.
The horizontal  dotted line indicates the radial velocity of the M86 nucleus.
\label{fig-vpt1ct1radpa}}
\end{figure}

The plot of the radial velocities 
  as a function of 
 the $(C-T_1)$ colors, $T_1$ magnitudes, and  
 position angles $\Theta$~(measured from the north to east)
 in Figure \ref{fig-vpt1ct1radpa} shows that 
the mean values of the radial velocities 
  do not change with either 
  the magnitude or the color. 
However,
  the mean values of the radial velocities 
  seem to depend on the position angle, showing a
  minimum value at $\Theta \approx 60^\circ$ and $\approx 300^\circ$.
  This suggests a rotation of the M86 GC system 
  (discussed in detail in the next section). 

\subsection{Rotation of the GC system}\label{rotation}

\begin{figure}
\epsscale{0.8}
\plotone{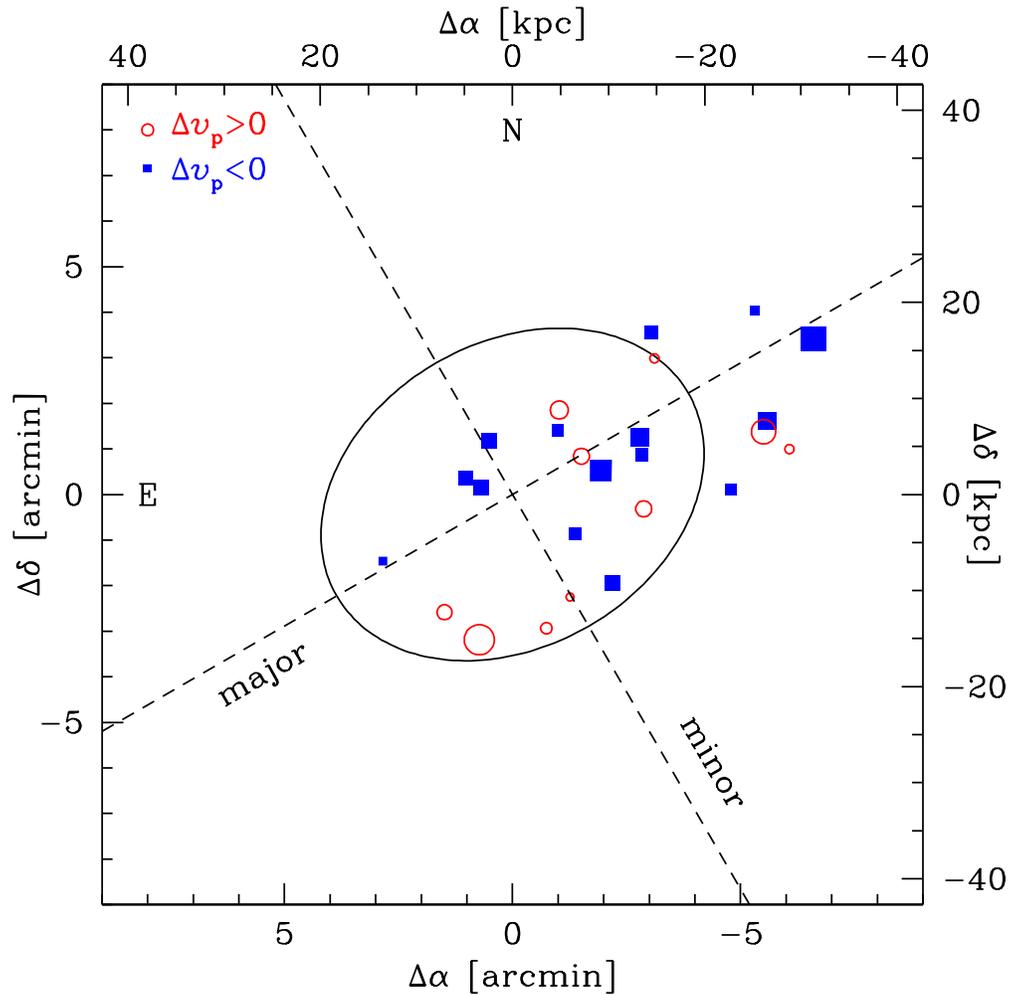}
\caption{ 
Spatial distribution of the 25 identified M86 GCs. 
GCs with velocities larger and smaller than
  the velocity 
  of the M86 nucleus are plotted
  by open circles and filled boxes, respectively.
The solid ellipse represents the boundary for the standard major diameter $D_{25}$ of M86 \citep{dev91}.
The dashed lines represent the photometric major and minor axes.   
\label{fig-radec}}
\end{figure}

In Figure \ref{fig-radec}, 
  we show the spatial distribution of the M86 GCs with measured velocities.
Although the spatial coverage is neither uniform nor large,
  the spatial segregation of the high- and low-velocity GCs 
  relative to the velocity of the M86 nucleus can be seen,
  which indicates a rotation of the GC system (see also Figure \ref{fig-vpt1ct1radpa} (c)).  

The amplitude and axis of rotation 
   for the M86 GC system was measured with the following assumptions:
  (a) the GC system is spherically symmetric,
  and (b) the rotation axis of the GC system lies in the plane of the sky.
If the GCs follow any overall rotation, 
 the radial velocities will depend sinusoidally on the azimuthal angles.
Thus, we can then determine the amplitude and axis of rotation
  by fitting the radial velocities ($v_p$) 
  to the function
  \citep{cot01,cot03,hwa08,lee10a},

\begin{equation}
v_p (\Theta) = v_{\rm sys} + (\Omega R) \sin(\Theta - \Theta_0)\ ,
\label{eq-rot}
\end{equation}

\noindent where  $\Omega R$ is the rotation amplitude 
  and $v_{\rm sys}$ is the systemic velocity of the GC system.
  
\begin{figure}
\epsscale{0.5}
\plotone{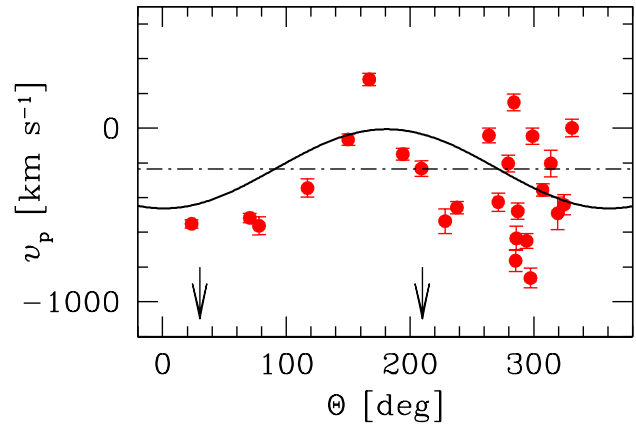}
\caption{ 
Radial velocity versus position angle of the M86 GCs. 
The solid curve is a  best fit rotation curve from  Table 4. 
The  horizontal dot-dashed line represents the velocity of the M86 nucleus.
The vertical arrows show the photometric minor axis of M86.
\label{fig-thetavp}}
\end{figure}

Figure \ref{fig-thetavp} plots the radial velocities of the GCs 
  as a function of position angle
and an overlay of the best fit rotation curve. 
We use an error-weighted, nonlinear fit of equation (\ref{eq-rot})
  with $v_{\rm sys}$ fixed to the value of the M86 nucleus velocity. 
This gives a rotation amplitude of $\Omega R$ $ = 228^{+71}_{-80}$ km s$^{-1}$, 
which suggests a rotation of the M86 GC system.
However, as this result is based on a small sample size, 
 it will need to be examined again with a larger number of GCs in future studies.

The orientation of the rotation axis ($\Theta_0$) is estimated to be
  ${91^\circ}^{+19}_{-21}$, 
 and this appears to be closer to the photometric major axis of M86 ($\Theta_{phot} = 120^\circ$)
  than to the minor axis.
The rotation of the M86 GC system around the major axis
  is consistent with the result based on stellar kinematics,
   although the spatial coverage for the stellar kinematics was 
  much smaller than this study \citep{kra11}.


We summarize the kinematic results of the M86 GC system in Table \ref{tab-m86kin}:
the range of galactocentric radius of the GCs in arcsec,
mean value of the radial distance in arcsec, number of GCs, mean projected velocity and
velocity dispersion about the mean velocity ($\sigma_{p}$), 
position angle of the rotation axis and rotation amplitude estimated using equation (\ref{eq-rot}), 
velocity dispersion about the best fit rotation curve ($\sigma_{p,r}$), 
and absolute value of the ratio of the rotation amplitude to the velocity dispersion about the best fit rotation curve.
%
The uncertainties of the values represent 
  $68\%$ $(1\sigma)$ confidence intervals that were
  determined from the numerical bootstrap procedure following the method of \citet{cot01}.

\subsection{Existence of Dark Matter Halo} 

Here, we investigate the existence of an extended dark matter halo in M86
  by a comparison of the observed velocity dispersion profile (VDP) of the GCs 
  and the VDP expected from the stellar mass profile
  \citep{cot01,cot03,hwa08,lee10a}.
The stellar mass profile is derived from the surface brightness profile of M86, which is then
used to compute the VDP. 
  
Assuming that the M86 GC system is spherically symmetric in the absence of rotation,
  we apply the spherical Jeans equation (e.g., \citealt{bin87})
  to a dynamical analysis of the GC system.
The spherical Jeans equation is 

\begin{equation}
{d\over{dr}}\, n_{\rm cl}(r) \sigma_r^2(r) + 
  {{2\,\beta_{\rm cl}(r)}\over{r}}\, n_{\rm cl}(r) \sigma_r^2(r) = - 
  n_{\rm cl}(r)\,{{G M_{\rm tot}(r)}\over{r^2}}\ , 
\label{eq-jeans}
\end{equation}

\noindent where $r$ is the three-dimensional radial distance from the galactic center,
  $n_{\rm cl}(r)$ is the three-dimensional density profile of the GC system,
  $\sigma_r(r)$ is the radial component of velocity dispersion,
  $\beta_{\rm cl}(r)\equiv1-\sigma_\theta^2(r)/\sigma_r^2(r)$ is the velocity anisotropy,
  $G$ is the gravitational constant,
  and $M_{\rm tot}(r)$ is the total gravitating mass contained
  within a sphere of radius $r$.
The tangential component of velocity dispersion, $\sigma_\theta(r)$,
  is equal to the azimuthal component of the velocity dispersion, $\sigma_\phi(r)$,
  in the absence of rotation. 
The total mass $M_{\rm tot}(r)$ interior out to any radius is 
  the sum of the dark matter mass $M_{\rm dm}(r)$ and stellar mass $M_{\rm s}(r)$.

To solve equation  (\ref{eq-jeans}),
  $n_{\rm cl}(r)$ and $M_{\rm s}(r)$ are obtained from the observation. 
If we fix $\beta_{\rm cl}(r)$ as a constant and assume $M_{\rm dm}(r)$,
  we can predict the profile of $\sigma_r(r)$.
A comparison of this 
   with the observed VDP, 
   will provide information about the existence of a dark matter halo.
Because the observed VDP for the GCs is projected, 
  we also need to compute the projected VDP, $\sigma_p(R)$: 
\begin{equation}
\sigma_p^2(R) = {2\over{N_{\rm cl}(R)}}\, \int_R^{\infty} n_{\rm
cl} \sigma_r^2(r) \left(1 - \beta_{\rm cl}\,{{R^2}\over{r^2}}
\right)\, {{r\,dr}\over{\sqrt{r^2 - R^2}}}, 
\label{eq-sigp}
\end{equation}
\noindent where $R$ is the projected galactocentric distance 
and the surface density profile, $N_{\rm cl}(R)$, is the
  projection of the three-dimensional density profile $n_{\rm cl}(r)$.

\subsubsection{Density Profile of the M86 GC System}

\begin{figure}
\epsscale{0.7}
\plotone{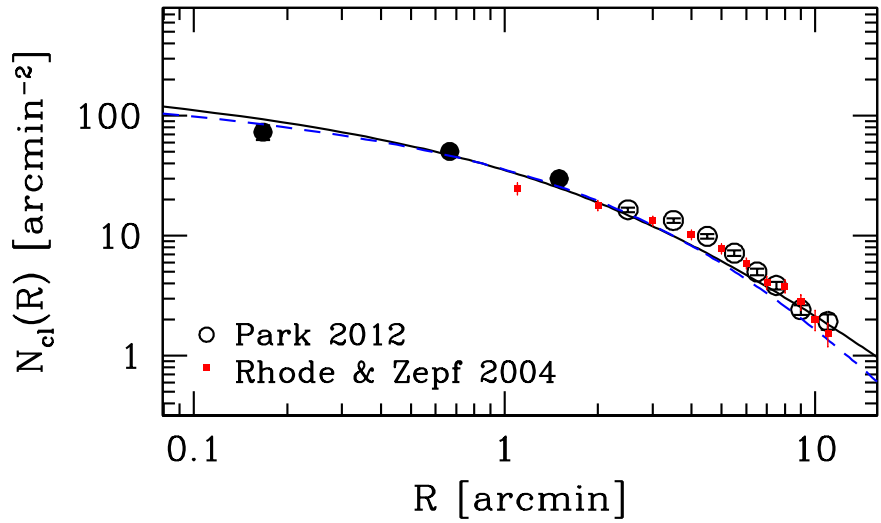}
\caption{ 
Projected surface number density profile for the M86 GCs. 
The filled and open circles represent the GC candidates from the HST/ACS images and
  those from the KPNO $CT_1$ images, respectively \citep{par12b}. 
  The small squares represent the results from \citet{rho04}.
The solid and dashed lines indicate the projected best fits using the NFW and Dehnen density profiles, respectively. 
\label{fig-numden}}
\end{figure}

The three-dimensional density profiles are derived from the surface number density
  of the M86 GCs using the Navarro-Frenk-White (NFW) profile \citep{nfw97} and the Dehnen profile \citep{deh93}. 
We adopt the surface density profile given in \citet{par12b},
  which was derived
  using data from the HST/ACS Virgo Cluster Survey \citep{jor09} 
  for GCs at $R<2$ arcmin and
  from the KPNO $CT_1$ data for GCs at $R>2$ arcmin.
The surface number density profile $N_{\rm cl}(R)$ 
  is shown in Figure \ref{fig-numden}.
We fit the surface number density profile 
  to the projections of the NFW profile, 
  $n_{\rm cl} (r)=n_0(r/b)^{-1}(1+r/b)^{-2}$ and 
  the Dehnen profile,
  $n_{\rm cl} (r)=n_0(r/a)^{-\Gamma}(1+r/a)^{\Gamma-4}$. 
 The profile $N_{\rm cl}(R)$ is
  derived from an integration of the three-dimensional density profile $n_{\rm cl}(r)$ as follows:
\begin{equation}
N_{\rm cl}(R) = 2\int_{R}^{\infty} n_{\rm cl}(r) {{r\,dr}\over{\sqrt{r^2-R^2}}}\ . 
\label{eq-ncl}
\end{equation}  
The results of the fit are summarized as follows: 
\begin{equation}
\begin{array}{rcl}
n_{\rm cl}^{\rm NFW}(r) & = &
  0.052\,{\rm kpc}^{-3}(r/14.58\,{\rm kpc})^{-1}(1+r/14.58\,{\rm kpc})^{-2} \\
n_{\rm cl}^{\rm Dehnen}(r) & = &  
  0.028\,{\rm kpc}^{-3}(r/27.94\,{\rm kpc})^{-0.93}(1+r/27.94\,{\rm kpc})^{-3.07}. \\ 
\end{array}
\label{eq-nfw}
\end{equation}
%

\subsubsection{Stellar Mass Profile}

\begin{figure}
\epsscale{1.0}
\plotone{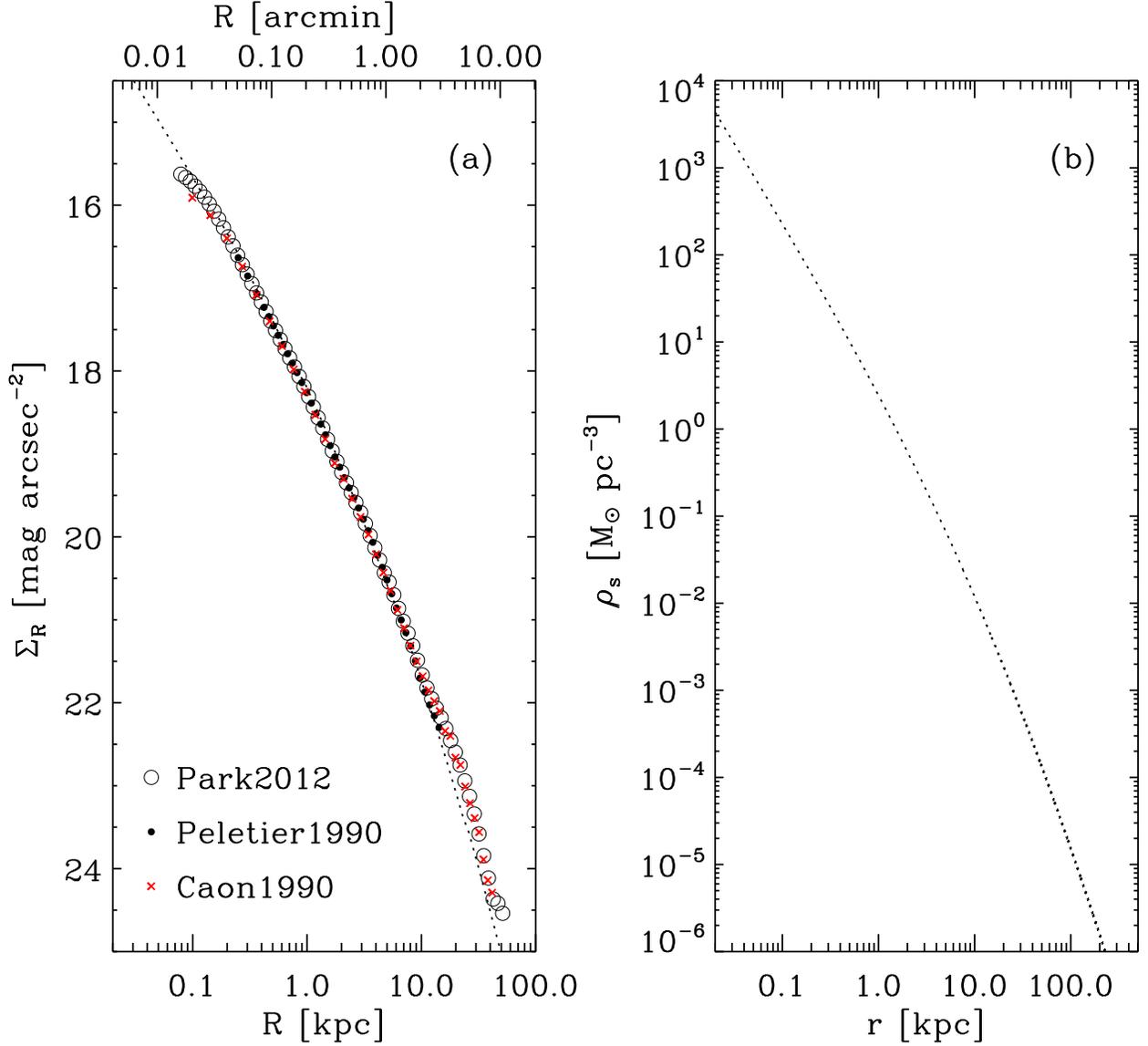}
\caption{ 
(a) Surface brightness profiles of M86.
The open and filled circles are $R$-band surface photometry given in \citet{par12b}
and in \citet{pel90}, respectively.
The crosses represent the surface photometry converted from $B$-band photometry given in \citet{cao90}.
The dotted line represents a projected best fit of the three-dimensional luminosity density profile.
(b) Three-dimensional stellar mass density profile
derived with a constant $R$-band mass-to-light ratio of $\Upsilon_0=6.5~M_\odot L^{-1}_{R,\odot}$.
\label{fig-stellsurf}}
\end{figure}

Figure \ref{fig-stellsurf} gives 
  the radial profile of the M86 surface brightness,
  derived from the KPNO $R$-band images \citep{par12b},
  and stellar mass.
This surface brightness profile agrees well 
  with the results in \citet{pel90} that are 
  based on $R$-band photometry and 
  with those in \citet{cao90} that are converted from $B$-band photometry.

A fit of the surface brightness profile of \citet{par12b} to 
  the projection of three-dimensional luminosity density profile 
  \citep{cot03, hwa08,lee10a}
  represented by
\begin{equation}
j(r) = {{(3-\gamma)(7-2\gamma)}\over{4}}\, {L_{\rm tot}\over{\pi a^3}}\, 
  \left({r\over{a}}\right)^{-\gamma}\,
  \left[1+\Bigl({r\over{a}}\Bigr)^{1/2} \right]^{2(\gamma-4)}\ ,
\label{eq-lumden}
\end{equation}
gives 
  $\gamma=1.75$, $L_{\rm tot}=1.60\times10^{11}~L_{R,\odot}$, and $a=29.94$ kpc.
  The projected best fit curve is overlaid in Figure \ref{fig-stellsurf} (a).
%

Figure \ref{fig-stellsurf} (b)
  shows a three-dimensional stellar mass density profile, $\rho_s(r)=\Upsilon_0 j(r)$,
  derived with an $R$-band mass-to-light ratio of $\Upsilon_0=6.5~M_\odot L^{-1}_{R,\odot}$
  (determined in the next section).
From this profile, 
  we obtain an M86 stellar mass profile represented by
\begin{eqnarray}
M_{\rm s}(r) & = &\int_{0}^{r}{4{\pi}x^2\,\rho_s(r)}dx  = \Upsilon_0\int_{0}^{r}{4{\pi}x^2\,j(x)}dx \nonumber \\
 & = & {\Upsilon_{0}}L_{\rm tot}\left[{(r/a)^{1/2}
    \over 1+(r/a)^{1/2}}\right]^{2(3-\gamma)}\left[{(7-2\gamma)+(r/a)^{1/2}
    \over 1+(r/a)^{1/2}}\right] \ .
\label{eq-starmass}
\end{eqnarray}

\subsubsection{Extended Dark Matter Halo in M86}

If we adopt the stellar mass profile as the total mass profile 
  (i.e., $M_{\rm tot}(r)=M_{\rm s}(r)$) and 
  $\rho_s(r)\propto j(r)$ instead of $n_{\rm cl}(r)$,  
  we can derive the radial component of VDPs
  for the stellar system from the Jeans equation.
We also assume $R$-band mass-to-light ratios ($\Upsilon_0$) and
  velocity anisotropies of the stellar halo ($\beta_{\rm s}(r)$).
The projected VDPs 
  are then computed 
  from the radial component of VDPs through equation (\ref{eq-sigp}).

\begin{figure}
\epsscale{1.0}
\plotone{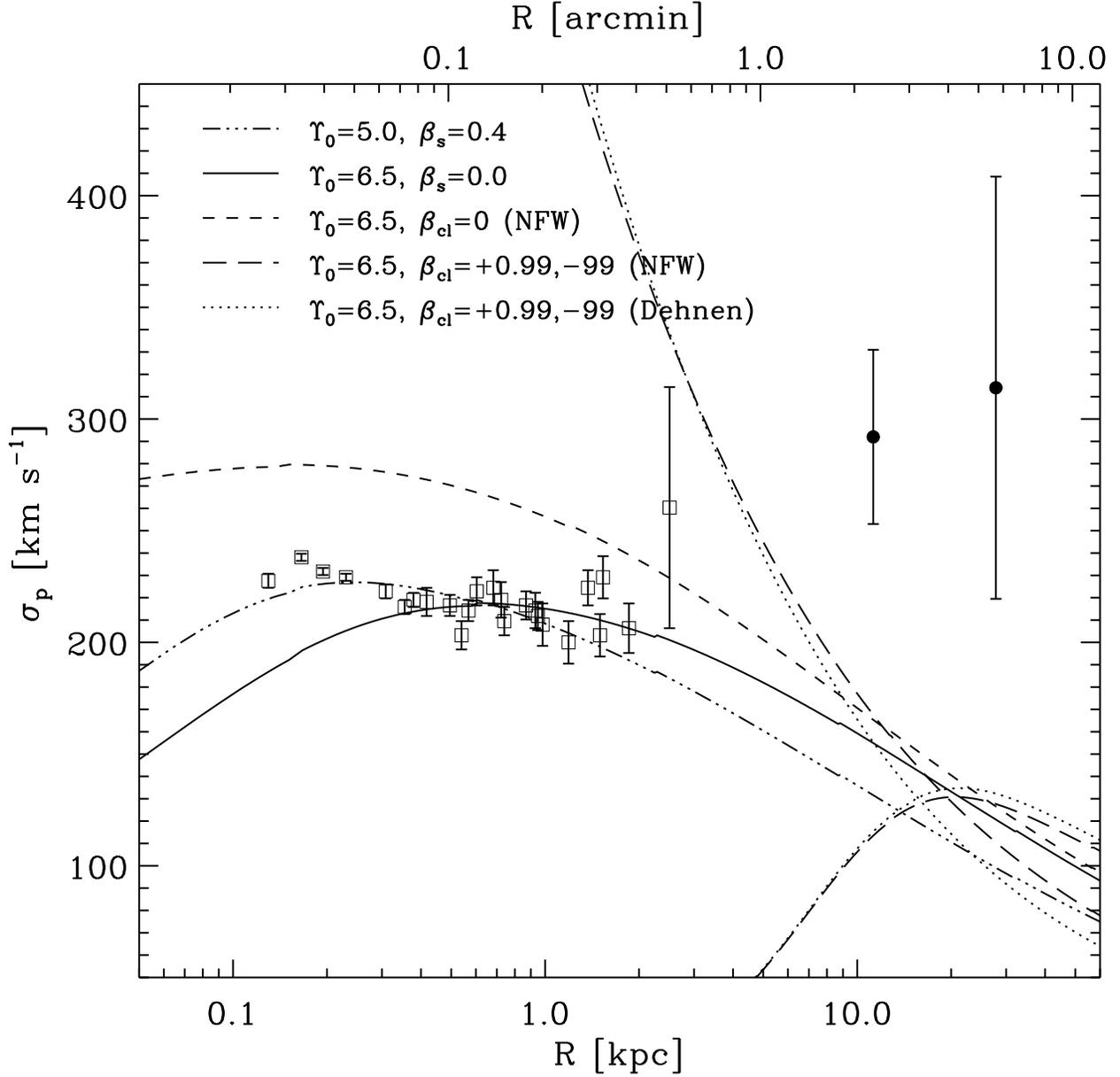}
\caption{ VDPs for the stars and  GCs in M86.
The open squares represent the stellar velocity dispersions from \citet{ben94}.
The filled circles show the velocity dispersions of the GCs.
The solid line represents the stellar VDP expected from 
  the stellar mass model with 
  a constant stellar mass-to-light ratio of $\Upsilon_0=6.5~M_\odot L^{-1}_{R,\odot}$
  and a stellar velocity anisotropy of $\beta_{\rm s}=0.0$.
The triple-dot-dashed curve shows  the stellar VDP 
 with $\Upsilon_0=5.0~M_\odot L^{-1}_{R,\odot}$ and $\beta_{\rm s}=0.4$.
Also shown are 
  VDPs expected from the same stellar mass model as above
  but with NFW GC density profiles and velocity anisotropies
  of $\beta_{\rm cl}=0.0$ (short dashed line) and $+0.99, -99$ (long dashed lines).
  The dotted lines represent the VDPs with a Dehnen density profile at $\beta_{\rm cl}=+0.99$ and $-99$. 
\label{fig-stargcvdp}}
\end{figure}

Figure \ref{fig-stargcvdp}
  plots the M86 GC VDP measured in this study
  and 
  the observed M86 stellar VDP given in \citet{ben94}. 
The stellar velocity dispersion is almost constant 
  around 220 km s$^{-1}$ in the inner region
 and is smoothly connected to the GC velocity dispersion at $R\approx 3$ kpc.
We also plot the predicted VDPs derived with 
  $\Upsilon_0=5.0~M_\odot L^{-1}_{R,\odot}$, $\beta_{\rm s}(r)=0.4$ (radially biased)
  and $\Upsilon_0=6.5~M_\odot L^{-1}_{R,\odot}$, $\beta_{\rm s}(r)=0.0$ (isotropic),
  which fit the observed stellar kinematic data at $R<2$ kpc well.
For comparison, the predicted VDPs for the M86 GCs 
  derived with the same stellar mass profile, 
  the GC number density profiles ($n_{\rm cl}^{\rm NFW}(r)$ and $n_{\rm cl}^{\rm Dehnen}(r)$) determined in the previous section, 
  and several velocity anisotropies ($\beta_{\rm cl}(r)=+$0.99 (radially biased),
  0.0 (isotropic),  and --99 (tangentially biased)) are also shown. 
The VDPs based on the NFW and Dehnen density profiles are not significantly different from each other. 
Note that none of the models agree with 
  the observed VDPs of the GCs at $R>3$ kpc,
  which suggests that the mass-to-light ratio is not constant over the galactocentric radius
  but should increase with radius.
This clearly suggests the existence of an extended dark matter halo in the outer region of M86.


\section{Metallicities and Ages} 

\subsection{Metallicity and Age Measurement}

We determined the metallicities and ages of the M86 GCs 
  using two methods:
  (1) the BH method, which determines the metallicity
  through an empirical relation 
  between absorption line index and metallicity developed by \citet{bro90},
  and 
  (2) the grid method, which derives the metallicity and age 
  from a comparison of the Lick line index of the spectrum 
  with that of a single stellar population (SSP) model.
We explain each method here. 

\subsubsection{BH Method}

\citet{bro90} and \citet{huc96} presented linear relations 
  between absorption line indices obtained from the integrated spectra
  and mean metallicities to derive the metal abundances of old stellar systems.
Their method was developed to minimize systematic effects such as 
  reddening, individual element abundance anomalies, and instrumental effects.
They recommend six indices (G band, MgH, Mg2, Fe5270, CNB, and $\Delta$) 
  as primary calibrators among the 12 line indices  
  for the empirical relations. 
Here, we use only four primary line indices 
  to determine the metallicities of the M86 GCs 
  (G band, MgH, Mg2, and Fe5270) 
  because of the low S/N of the spectra in the wavelength range 
  for the  CNB and $\Delta$ indices.

Each spectrum is first shifted into the rest frame, and
we then measure the absorption line indices 
  following the prescription of \citet{bro90} and \citet{huc96}.
The measured absorption line indices are calibrated to the BH index system
  with a zero point offset, $Index({\rm BH})=Index({\rm Subaru})+ offset$, 
  determined from the indices of five MW GCs that are common to this study and \citet{huc96}.
The offsets we derived are
  $0.014 \pm 0.015$ for G band, 
  $-0.013 \pm 0.009$ for MgH, 
  $-0.021 \pm 0.018$ for Mg2, 
  and $0.008 \pm 0.009$ for Fe5270. 
We determine the metallicity from each index
  as follows:
  $\textrm{[Fe/H]}_{G band}=11.415\times \textrm{Gband}-2.455$,
  $\textrm{[Fe/H]}_{MgH}=20.578 \times \textrm{MgH}-1.840$,
  $\textrm{[Fe/H]}_{Mg2}=9.921 \times \textrm{Mg2}-2.212$, and
  $\textrm{[Fe/H]}_{Fe5270}=20.367 \times \textrm{Fe5270}-2.086$.
Finally,
  we take an error-weighted average of the four measurements
  as the final metallicity value for each GC.
Here, the error 
  is the mean of the standard deviation.
We are able to determine the metallicity for 16 of 25 GCs 
  based on this method.
The metallicities for the other nine GCs could not be determined because of the low S/N of the spectra.

\subsubsection{Lick Index Grid Method}

Lick absorption line indices are useful for determining 
  the metallicity and age of old stellar systems
  from a comparison of 
  the indices derived from the spectra 
  with the line index grids 
  predicted from SSP models \citep{tri95, tra00, tho04, puz05, bea08, tra08, woo10}. 
For this grid method,
 we adopt the SSP models given by \citet{tho03,  tho04, tho05} and
 follow the technique described in \citet{puz05} and \citet{par12a}.
  
  We calibrate our absorption line indices to the Lick index system as follows.
The spectra are smoothed with the Lick resolution \citep{wor97}
  after shifting each spectrum into the rest frame.
Lick line indices are then derived from the spectra of the M86 GCs,
  following the definitions given in \citet{wor94} and \citet{wor97}.
Here, the line index errors are derived from the photon noise in the spectra
  before the flux calibration.
The resulting line indices are then calibrated 
  to the Lick system with the zero point offset,
  $Index({\rm Lick})=Index({\rm Subaru})+offset$, 
  determined from the spectra of five MW GCs that are common to this study,
  \citet{tra98}, and \citet{kun02}.
The offsets we derived are
  $0.184 \pm 0.140$ for \hbetaa, 
  $0.165 \pm 0.187$ for Mgb, 
  $0.443 \pm 0.306$ for Fe5270, 
  and $-0.103 \pm 0.162$ for Fe5335
  (see Table 1 in \citet{par12a} for the offsets of other indices).
We determined the Lick line indices for eight GCs in M86,
  which are listed with errors in Table \ref{tab-lickindexerr}.

\begin{figure}
\epsscale{0.5}
\plotone{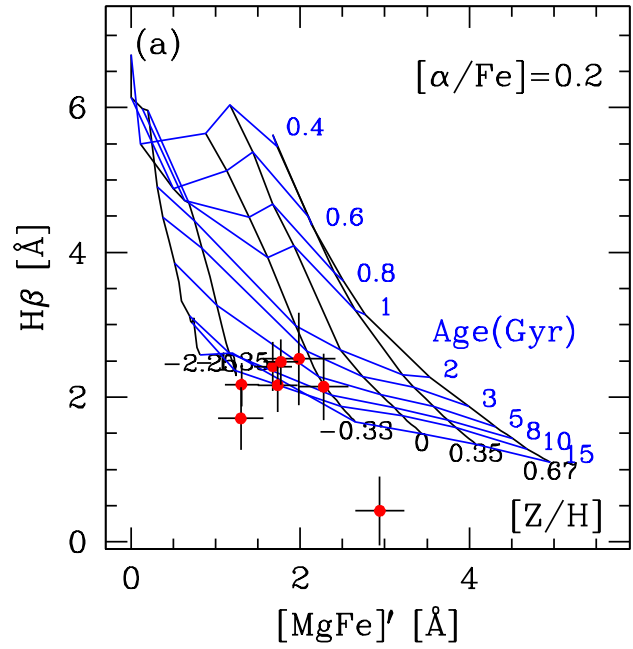}
\caption{ 
\hbeta versus \mgfep for the M86 GCs.
The grids represent the SSP models  with \afe= 0.2
  for various values of [Z/H] (--2.25, --1.35, --0.33, 0, 0.35, and 0.67) and ages (0.4 -- 15 Gyr) given by \citet{tho03}. 
\label{fig-samplegrid}}
\end{figure}

The composite index \mgfepp, 
  defined by
  \mgfep$= \sqrt{\rm{Mgb} \times\ (0.72 \times \rm{Fe5270} + 0.28 \times \rm{Fe5335})}$,
  is a good metallicity tracer because of its low sensitivity to \afee.
The index \hbeta is an age indicator and
  the least sensitive to \afe among the Balmer lines \citep{tho03}.
Thus, we determine the metallicity and age of each GC
  in the \hbeta versus \mgfep grids 
  provided by \citet{tho03}.
Figure \ref{fig-samplegrid} shows 
   the observational indices of the M86 GCs 
   in comparison with the SSP model grids for \hbeta versus \mgfepp,
  which indicate  SSP models with  \afe = 0.2, 
  [Z/H] = --2.25, --1.35, --0.33, 0.0, 0.35, and 0.67 dex, and 
   ages of 0.4, 0.6, 0.8, 1, 2, 3, 5, 8, 10, and 15 Gyr.
All GCs seem to have ages larger than $\sim$5 Gyr 
  and metallicities smaller than the solar abundance.   
For the GCs inside the envelope of the model grid,
  we take the [Z/H] value and age at 
  the nearest model grid interpolated with bins of 0.01 dex 
  for [Z/H] and 0.1 Gyr for age.
For the GCs outside the envelope,  
  we take the values of 
  the nearest envelope of the model grid 
  in the direction of the error vector as done in \citet{puz05}.
The two outliers with large difference from the model envelope along the \hbeta axis
  might be due to a limit of the low resolution integrated spectroscopy or of the model grids
  because these objects are also often shown in studies of MW GCs and M31 GCs 
 derived from high S/N spectra \citep{puz02,puz05,sch12} as well as other gE GCs \citep{cen07,woo10}.
To estimate the errors in the age and metallicity,
  we calculate the ages and metallicities of four data points
  composed of \hbeta $\pm$ error and \mgfep $\pm$ error in the grid. 
The difference between the average of these four values and 
  the estimate calculated
  directly from the $index$ is taken as the final error.

\begin{figure}
\epsscale{1.0}
\plotone{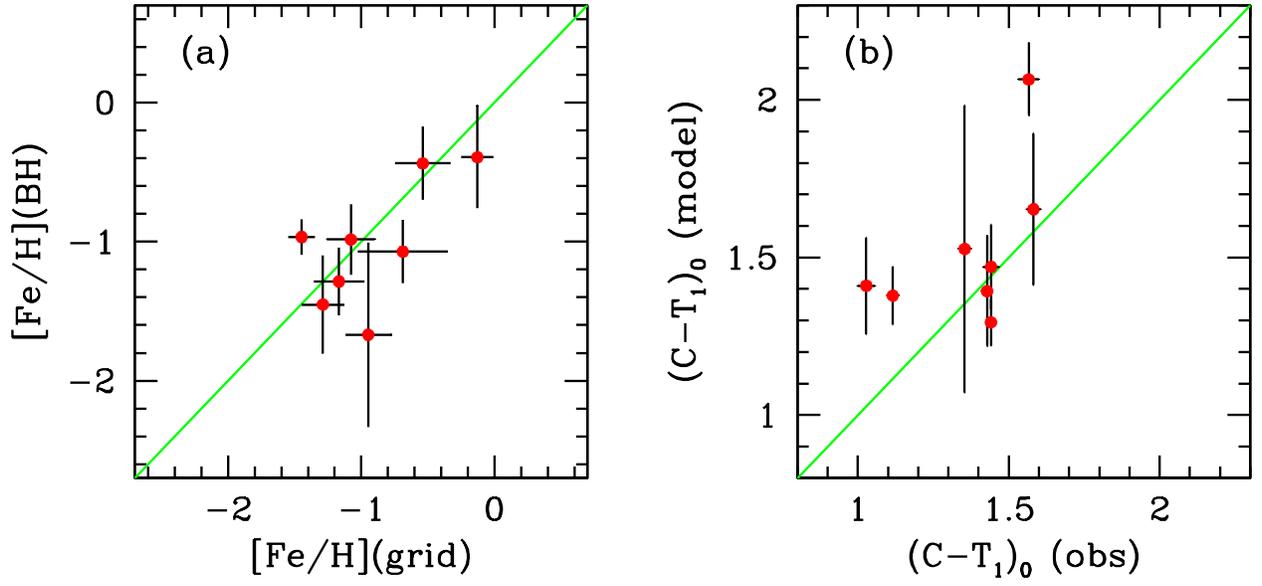}
\caption{ 
(a) Comparison of the M86 GC metallicities measured using the BH and grid methods.
(b) Comparison of the photometric colors obtained from the observation and a model.
$(C-T_1)_0$ (obs) are the observational colors from \citet{par12b} and 
$(C-T_1)_0$ (model) are the model colors derived from the [Z/H] and age of the M86 GCs 
using the SSP model of \citet{mar08}.  
The solid lines represent one-to-one relations.
%
%
\label{fig-compBHgrid}}
\end{figure}

The metallicities of eight GCs derived from the grid method
  are compared with those from the BH method in Figure \ref{fig-compBHgrid} (a). 
  The two measurements are broadly consistent within the uncertainty.
Here, the total metallicity ([Z/H]) derived from the grid method
  was converted into [Fe/H](grid) using the relation 
  $\textrm{[Fe/H]}=\textrm{[Z/H]}-0.94$ \afe 
  \citep{tho03}.
For this conversion, we adopted \afe = 0.2, 
  which is the mean \afe for the GCs in gEs \citep{par12a}.
We also compared the observational $(C-T_1)_0$  colors of the M86 GCs \citep{par12b} with
 the model colors derived from the [Z/H] and age 
 using the SSP model in \citet{mar08}.
Among the eight GCs, five GCs agree well within their errors
  as shown in Figure \ref{fig-compBHgrid} (b).
%

\subsection{Metallicities and Ages of the M86 GCs}

\begin{figure}
\epsscale{0.8}
\plotone{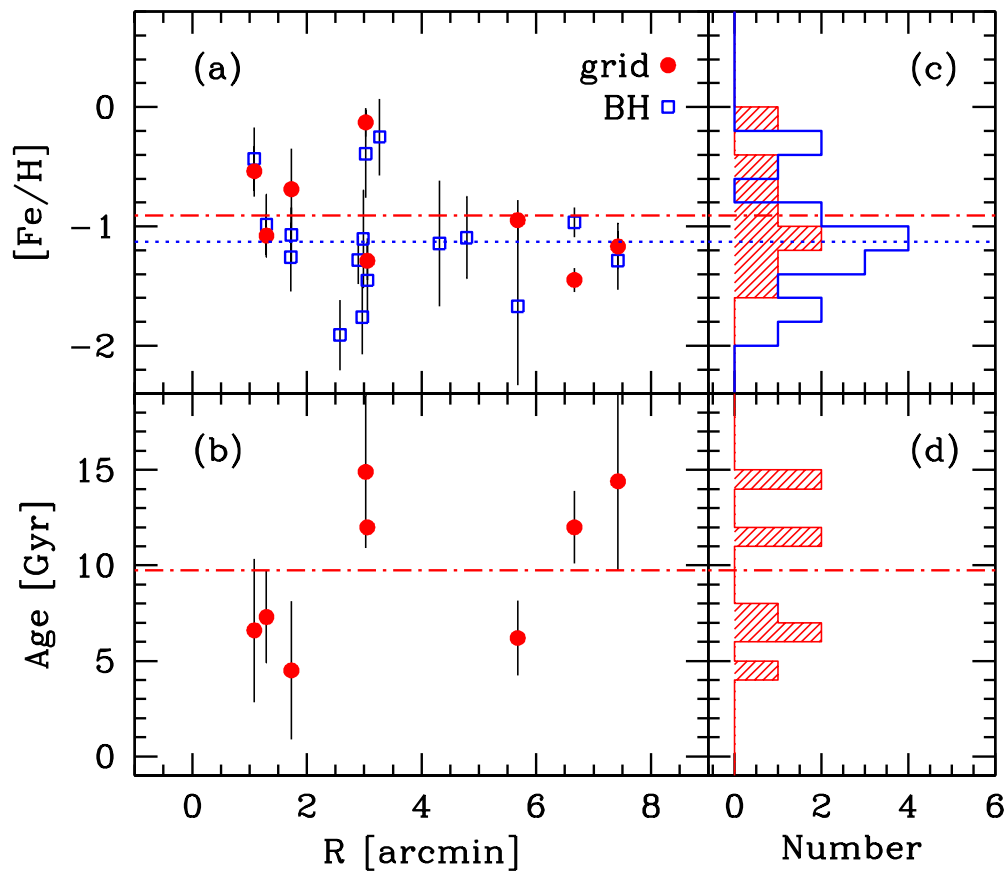}
\caption{ 
Metallicities and ages of the M86 GCs as a function of galactocentric radii,
  and their histograms.
The filled circles and  hatched histograms indicate the values
  from the grid method, while
  the open squares and solid histogram are those from the BH method.
The horizontal dot-dashed and dotted lines 
  indicate the mean values of each parameter
  measured from the grid and BH methods, respectively.
\label{fig-agemetal}}
\end{figure}

The metallicities and ages of the M86 GCs are listed in Table \ref{tab-m86gc}, and 
Figure \ref{fig-agemetal} shows 
(a) [Fe/H] versus $R$, (b) age versus $R$, (c) [Fe/H] distribution,
and (d) age distribution of the M86 GCs. 
It is not easy to derive any systematic trend in the data because of small number statistics.
However, several features are noted.
First, the metallicities of the 16 GCs based on the BH method show a wide range of $-2.0 <$ [Fe/H] $<-0.2$
  with a mean value of  $-1.13 \pm 0.47$,
  which is similar to that based on the the grid method  ($-0.91 \pm 0.44$).
Second, metal-rich GCs ([Fe/H] $>-0.9$) are found only in the inner region ($R \leq 4\arcmin $).
Third, the age of the M86 GCs also shows a wide range from 4 to 15 Gyr with a
mean of 9.7 $\pm$ 4.0 Gyr. This is similar to the GCs in other gEs  (e.g., \citealt{woo10,par12a}). 

\begin{figure}
\epsscale{0.5}
\plotone{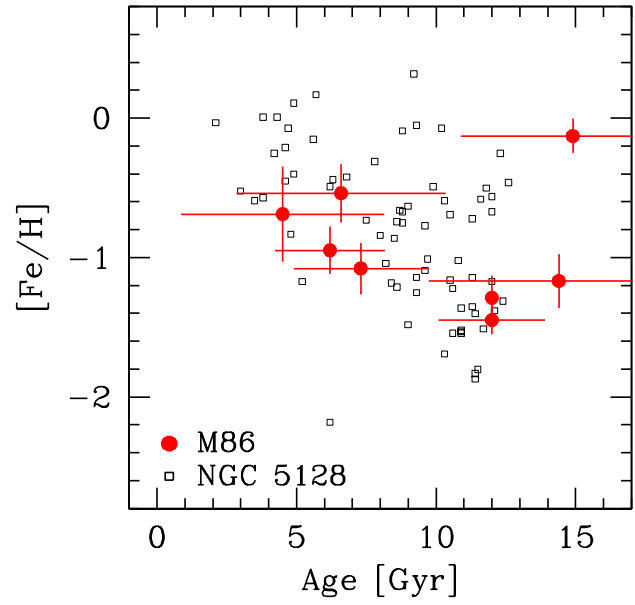}
\caption{ 
Age versus metallicity of the GCs in M86.
The circles and squares represent the GCs of M86 and NGC 5128, respectively.
%
\label{fig-gEsagemetal}}
\end{figure}

Figure  \ref{fig-gEsagemetal}  shows the metallicity as a function of age for the M86 GCs. 
  This figure shows a hint
  for an age-metallicity relation, meaning that the younger GCs are more metal-rich.
This relationship in gEs has also been seen for the GCs in M60 \citep{pie06} and in NGC 5128 \citep{woo10}.  
However, \citet{woo10} did not strongly conclude its existence due to the large biases in their selected GC sample (e.g., extremely bright clusters).
A more complete analysis with a more comprehensive
data set of gE GCs is necessary to draw a strong conclusion about the
age-metallicity relation.
    

\section{Summary}

Using the Subaru spectroscopic data of M86 GCs,
  we studied the kinematic and chemical properties of the M86 GC system.
Our main results are summarized as follows. 

\begin{enumerate}

 \item For the first time,
  we measured the radial velocities of 31 objects in the M86 field:
  25 M86 GCs, 
  two foreground stars, one probable intracluster GC in the Virgo cluster, 
  two faint galaxies, and the M86 nucleus.
  
 \item The mean velocity of the GCs is $\overline{v_p}=-354^{+81}_{-79}$ \kmss, which is
   different from the velocity of the M86 nucleus ($v_{\rm gal}=-234\pm 41$ \kmss). 
The velocity dispersion of the GCs is $\sigma_p=292^{+32}_{-32}$ \kmss.
The M86 GC system shows a hint of rotation.

\item From a comparison of the VDPs predicted from the stellar mass profile
  with the observed VDPs of the stars and GCs,
  we found evidence for the existence of an extended dark matter halo in M86.

\item  
We determined the metallicities for 16 GCs using the BH method, 
    and the ages and metallicities for 8 GCs using the grid method.
The metallicity of the M86 GCs derived from the BH method 
  is in the range $-2.0<$ [Fe/H] $<-0.2$  with a mean value of $-1.13 \pm 0.47$. 
The grid method results in similar [Fe/H] values 
  and a mean age of 9.7 $\pm$ 4.0 Gyr.
   

\end{enumerate}

\acknowledgments 
The authors like to thank the anonymous referee for his/her useful comments that improved the original manuscript.
The authors are grateful to the staff of the SUBARU Telescope
 for their kind help during the observation 
 and to our collaborator Nobuo Arimoto.
This is supported in part
by the Mid-career Researcher Program through an NRF grant funded by the MEST (No.2010-0013875).
H.S.H. acknowledges the support of the Smithsonian Institution.

\clearpage


\begin{deluxetable}{ccccccccccc}
\tabletypesize{\scriptsize}
\tablewidth{0pc} 
\tablecaption{Basic Properties of M86\label{tab-m86gal}}
\tablehead{ 
\colhead{Galaxy} & \colhead{$M_V$\tablenotemark{~a}} &
\colhead{$\upsilon_{\rm sys}$\tablenotemark{~b}} &
\colhead{R$_{\rm eff}$\tablenotemark{~c}} &
\colhead{$\epsilon$\tablenotemark{~d}} &
\colhead{P.A.$_{\rm min}$\tablenotemark{~e}} &
\colhead{Distance\tablenotemark{~f}} &
\colhead{$\sigma_{star}$\tablenotemark{~g}} &
\colhead{log($L_X$)\tablenotemark{~h}} &
\colhead{N$_{GC}$\tablenotemark{~i}} &
\colhead{S$_{N}$\tablenotemark{~j}} \\
\colhead{} & \colhead{} & \colhead{(km s$^{-1}$)} &
\colhead{(kpc)} & & \colhead{(deg)} & \colhead{(Mpc)} &
\colhead{(km s$^{-1}$)} & \colhead{(erg s$^{-1}$)} &
\colhead{blue red} & \colhead{} 
}
\startdata
M86      & --22.7 & --234 & 15.36 & 0.33 &  29 & 16.86 & 259$\pm$28 & 42.00$\pm$0.002 & 1453 968 &  3.5$\pm$0.5 \\
\enddata
\tablenotetext{\it{a}~} {
 $V$-band absolute total magnitude: \citet{kor09}. 
$^b$ Systemic velocity: this study.
$^c$ Effective radius in $R$-band: \citet{par12b}. 
$^d$ Ellipticity at R$_{\rm eff}$: \citet{par12b}.
$^e$ Position angle of the minor axis at R$_{\rm eff}$: \citet{par12b}.
$^f$ Distance: \citet{mei07}. 
$^g$ Mean velocity stellar dispersion at $R\lesssim 45\arcsec$: \citet{ben94}. 
$^h$ {Logarithmic} value of X-ray luminosity: \citet{beu99}.
$^i$ Numbers of blue GCs and red GCs: \citet{rho04}.
$^j$ Specific frequency of GCs: \citet{rho04}.
  }
\end{deluxetable}


\begin{deluxetable}{clccccccc}
\tablewidth{0pc}
\tablecaption{Observing Log for the Subaru FOCAS/MOS Run\label{tab-mask}}
\tablehead{
\colhead{Mask Name} &
\colhead{R.A.} &
\colhead{Decl.} &
\colhead{N(objects)} &
\colhead{T(exp)} &
\colhead{seeing} &
\colhead{Date} \\
\colhead{} &
\colhead{(J2000)} &
\colhead{(J2000)} &
\colhead{} &
\colhead{(s)} &
\colhead{(\arcsec)} &
\colhead{(UT)} 
}
\startdata
Mask-C & 12:26:11.7 & 12:56:17 & 34 & 3$\times$1200  & 0.6 & Apr 21, 2002 \\
Mask-1 & 12:25:52.3 & 12:59:39 & 33 & 1$\times$1800  & 0.6 & Apr 21, 2002 \\
\enddata
\end{deluxetable}

\begin{deluxetable}{rrrrrrrrrrrc}
\rotate
\tabletypesize{\tiny} 
\tablewidth{0pc}
\tablecaption{Radial Velocity, Age, and Metallicity of Globular Clusters in M86 \label{tab-m86gc}}
\tablehead{
\colhead{ID$^a$} &
\colhead{R.A.} &
\colhead{Decl.} &
\colhead{$R$} &
\colhead{$\Theta$} &
\colhead{$T_1$} &
\colhead{$(C-T_1)$} &
\colhead{$v_p$} &
\colhead{Age} &
\colhead{[Z/H]} &
\colhead{[Fe/H]$_{BH}$} & \colhead{Mask} \\
\colhead{} &
\colhead{(J2000)} &
\colhead{(J2000)} &
\colhead{(arcsec)} &
\colhead{(deg)} &
\colhead{(mag)} &
\colhead{(mag)} &
\colhead{(km s$^{-1}$)} &
\colhead{(Gyr)} &
\colhead{(dex)} &
\colhead{(dex)} & \colhead{}
}
\startdata
\multicolumn{11}{c}{}\\ \multicolumn{11}{c}{\underbar{Globular Clusters}}\\ \multicolumn{11}{c}{}\\
   448 & 12:26:07.53 & 12:58:37.6 & 127.1 & 331.0 & $21.31\pm0.02$ & $1.41\pm0.03$ & $    2\pm 51$ & $     ...     $ & $ ...           $ & $      ...      $ & C\\
   284 & 12:26:07.63 & 12:58:10.4 & 103.3 & 324.4 & $20.88\pm0.02$ & $1.28\pm0.02$ & $ -441\pm 59$ & $     ...     $ & $ ...           $ & $ -1.26\pm  0.29$ & C\\
    65 & 12:26:13.87 & 12:57:57.7 &  77.7 &  23.5 & $19.72\pm0.01$ & $1.49\pm0.02$ & $ -551\pm 24$ & $  7.3\pm  2.4$ & $ -0.89\pm  0.18$ & $ -0.98\pm  0.25$ & C\\
   270 & 12:26:05.54 & 12:57:36.8 & 103.8 & 299.0 & $20.84\pm0.01$ & $1.41\pm0.02$ & $  -46\pm 47$ & $  4.5\pm  3.6$ & $ -0.50\pm  0.34$ & $ -1.07\pm  0.23$ & C\\
   316 & 12:26:03.79 & 12:57:18.1 & 120.6 & 285.3 & $20.98\pm0.01$ & $1.32\pm0.02$ & $ -763\pm 64$ & $     ...     $ & $ ...           $ & $      ...      $ & C\\
   107 & 12:26:15.94 & 12:57:08.0 &  65.0 &  70.6 & $20.09\pm0.01$ & $1.64\pm0.02$ & $ -517\pm 26$ & $  6.6\pm  3.8$ & $ -0.35\pm  0.21$ & $ -0.44\pm  0.26$ & C\\
   430 & 12:26:14.59 & 12:56:55.3 &  42.4 &  77.9 & $21.31\pm0.02$ & $1.91\pm0.05$ & $ -563\pm 51$ & $     ...     $ & $ ...           $ & $      ...      $ & C\\
   265 & 12:25:59.94 & 12:56:27.6 & 173.6 & 263.8 & $20.82\pm0.02$ & $1.38\pm0.03$ & $  -42\pm 43$ & $     ...     $ & $ ...           $ & $ -1.28\pm  0.20$ & C\\
   324 & 12:26:06.09 & 12:55:54.4 &  97.7 & 237.8 & $21.02\pm0.02$ & $1.62\pm0.04$ & $ -457\pm 36$ & $     ...     $ & $ ...           $ & $      ...      $ & C\\
   289 & 12:26:02.72 & 12:54:49.3 & 176.4 & 228.4 & $20.91\pm0.01$ & $1.76\pm0.03$ & $ -536\pm 71$ & $     ...     $ & $ ...           $ & $      ...      $ & C\\
   332 & 12:26:06.58 & 12:54:31.5 & 154.6 & 209.3 & $21.05\pm0.01$ & $1.37\pm0.03$ & $ -233\pm 45$ & $     ...     $ & $ ...           $ & $ -1.91\pm  0.29$ & C\\
   143 & 12:26:08.69 & 12:53:50.4 & 181.6 & 194.2 & $20.31\pm0.02$ & $1.62\pm0.03$ & $ -150\pm 33$ & $ 14.9\pm  4.0$ & $  0.06\pm  0.12$ & $ -0.39\pm  0.37$ & C\\
    77 & 12:26:14.73 & 12:53:35.2 & 196.2 & 167.1 & $19.85\pm0.01$ & $1.67\pm0.02$ & $  282\pm 36$ & $     ...     $ & $ ...           $ & $ -0.25\pm  0.32$ & C\\
   413 & 12:26:23.40 & 12:55:18.5 & 191.7 & 117.3 & $21.26\pm0.02$ & $1.66\pm0.03$ & $ -345\pm 53$ & $     ...     $ & $ ...           $ & $      ...      $ & C\\
   299 & 12:26:17.86 & 12:54:11.4 & 178.9 & 150.0 & $20.93\pm0.01$ & $1.64\pm0.03$ & $  -66\pm 32$ & $     ...     $ & $ ...           $ & $ -1.11\pm  0.41$ & C\\
    58 & 12:25:49.93 & 13:00:48.0 & 399.9 & 307.2 & $19.66\pm0.01$ & $1.50\pm0.01$ & $ -356\pm 36$ & $ 12.0\pm  1.9$ & $ -1.26\pm  0.10$ & $ -0.97\pm  0.13$ & 1 \\
    79 & 12:25:44.66 & 13:00:11.0 & 445.6 & 297.3 & $19.86\pm0.02$ & $1.50\pm0.03$ & $ -864\pm 57$ & $ 14.4\pm  4.7$ & $ -0.98\pm  0.19$ & $ -1.29\pm  0.25$ & 1 \\
   302 & 12:25:48.78 & 12:58:22.6 & 349.3 & 286.0 & $20.93\pm0.01$ & $1.35\pm0.02$ & $ -635\pm 70$ & $     ...     $ & $ ...           $ & $      ...      $ & 1 \\
    96 & 12:25:49.15 & 12:58:08.9 & 340.5 & 284.0 & $20.00\pm0.01$ & $1.09\pm0.03$ & $  149\pm 49$ & $  6.2\pm  2.0$ & $ -0.76\pm  0.17$ & $ -1.67\pm  0.66$ & 1 \\
   251 & 12:25:46.82 & 12:57:46.1 & 369.2 & 279.3 & $20.78\pm0.02$ & $1.50\pm0.04$ & $ -204\pm 49$ & $     ...     $ & $ ...           $ & $      ...      $ & 1 \\
   352 & 12:25:52.11 & 12:56:52.7 & 287.2 & 271.3 & $21.10\pm0.01$ & $1.43\pm0.04$ & $ -426\pm 52$ & $     ...     $ & $ ...           $ & $ -1.09\pm  0.34$ & 1 \\
   258 & 12:25:59.27 & 13:00:19.8 & 280.7 & 319.5 & $20.79\pm0.03$ & $1.27\pm0.04$ & $ -490\pm 93$ & $     ...     $ & $ ...           $ & $      ...      $ & 1 \\
   142 & 12:25:58.97 & 12:59:45.7 & 258.9 & 313.8 & $20.30\pm0.02$ & $1.44\pm0.03$ & $ -203\pm 77$ & $     ...     $ & $ ...           $ & $ -1.14\pm  0.53$ & 1 \\
   149 & 12:26:00.32 & 12:58:01.6 & 183.2 & 294.2 & $20.34\pm0.01$ & $1.18\pm0.02$ & $ -649\pm 43$ & $ 12.0\pm  0.1$ & $ -1.10\pm  0.16$ & $ -1.45\pm  0.35$ & 1 \\
   150 & 12:26:00.13 & 12:57:38.8 & 177.8 & 287.1 & $20.34\pm0.01$ & $1.29\pm0.02$ & $ -479\pm 47$ & $     ...     $ & $ ...           $ & $ -1.76\pm  0.31$ & 1 \\
\multicolumn{11}{c}{}\\ \multicolumn{11}{c}{\underbar{Stars}}\\ \multicolumn{11}{c}{}\\
    81 & 12:26:07.30 & 12:55:21.1 & 107.3 & 217.3 & $19.89\pm0.02$ & $1.87\pm0.03$ & $   10\pm 27$ & $  ... $ & $ ... $ & $  ... $ & C \\
   226 & 12:25:44.08 & 12:59:40.6 & 440.3 & 293.3 & $20.73\pm0.01$ & $2.81\pm0.04$ & $   81\pm 44$ & $     ...     $ & $ ...           $ & $      ...      $ & 1 \\
\multicolumn{11}{c}{}\\ \multicolumn{11}{c}{\underbar{Probable Intracluster Globular Cluster in the Virgo Cluster}}\\ \multicolumn{11}{c}{}\\
   446 & 12:26:04.79 & 12:54:10.2 & 186.4 & 213.1 & $21.30\pm0.03$ & $0.91\pm0.04$ & $ 2434\pm 52$ & $     ...     $ & $ ...           $ & $      ...      $ & C \\
\multicolumn{11}{c}{}\\ \multicolumn{11}{c}{\underbar{Galaxies}}\\ \multicolumn{11}{c}{}\\
     VCC 0833 & 12:25:44.63 & 13:01:19.3 & 481.2 & 304.5 & ... & ... & $  777\pm 38$ & $ 14.0\pm  0.0$ & $ -0.64\pm  0.09$ & $ -0.90\pm  0.31$ & 1 \\
     NGC 4406B & 12:26:15.18 & 12:57:49.8 &  80.9 &  38.4 & ... & ... & $  949\pm 23$ & $ 14.0\pm  0.1$ & $ -0.58\pm  0.12$ & $ -1.12\pm  0.39$ & C \\
    M86 & 12:26:11.74 & 12:56:46.4 &   0.0 &   0.0 & ... & ... & $ -234\pm 41$ & $  4.0\pm  0.3$ & $  0.26\pm  0.02$ & $  0.62\pm  0.46$ & C \\
\enddata
\tablenotetext{a~}{From \citet{par12b}.}
\end{deluxetable}
\clearpage

\begin{deluxetable}{cccrrrrrr}
\tablewidth{0pc} 
\tablecaption{Kinematics of the M86 Globular Cluster System\label{tab-m86kin}}
\tablehead{
\colhead{$R$} & \colhead{$\langle R\rangle$} & \colhead{$N$} &
\colhead{$\overline{v_p}$} & \colhead{$\sigma_p$} 
& \colhead{$\Theta_0$} & \colhead{$\Omega R$} 
& \colhead{${\sigma}_{p,r}$ 
} 
& \colhead{$\Omega R$/${\sigma}_{p,r}$} \\
\colhead{(arcsec)} & \colhead{(arcsec)} & \colhead{} & 
\colhead{(km s$^{-1}$)} & \colhead{(km s$^{-1}$)} & \colhead{(deg)} & \colhead{(km s$^{-1}$)} & 
\colhead{(km s$^{-1}$)} & \colhead{}  
}
\startdata
42 $-$ 446 & 203 & 25 & $-354^{+81}_{-79}~$ & $292^{+32}_{-32}~$ & $91^{+19}_{-21}~$ & $228^{+71}_{-80}~$ & $282^{+36}_{-33}~$ & $0.81^{+0.32}_{-0.30}$ \\ 
\enddata
\end{deluxetable}

\begin{deluxetable}{crrrrr} 
\tablewidth{0pc} 
\tablecaption{Lick line indices and errors\label{tab-lickindexerr}}
\tablehead{ 
\colhead{ID} & \colhead{H$\beta$} & \colhead{Mg2}   & \colhead{Mgb}   & \colhead{Fe5270} & \colhead{Fe5335} \\
\colhead{ }  & \colhead{(\AA)}    & \colhead{(mag)} & \colhead{(\AA)} & \colhead{(\AA)}  & \colhead{(\AA)} }
\startdata
     65 &   2.423 $ \pm$   0.340 &   0.120 $ \pm$   0.006 &   2.765 $ \pm$   0.340 &   0.996 $ \pm$   0.359 
&   1.074 $ \pm$   0.459 \\  
    270 &   2.527 $ \pm$   0.641 &   0.075 $ \pm$   0.012 &   3.316 $ \pm$   0.623 &   2.496 $ \pm$   0.659 
&  --2.172 $ \pm$   0.909 \\  
    107 &   2.142 $ \pm$   0.458 &   0.179 $ \pm$   0.008 &   3.635 $ \pm$   0.444 &   2.037 $ \pm$   0.457 
&  --0.147 $ \pm$   0.602 \\  
    143 &   0.427 $ \pm$   0.474 &   0.102 $ \pm$   0.008 &   3.255 $ \pm$   0.448 &   2.762 $ \pm$   0.478 
&   2.387 $ \pm$   0.574 \\  
     58 &   2.173 $ \pm$   0.301 &   0.109 $ \pm$   0.005 &   0.814 $ \pm$   0.307 &   2.114 $ \pm$   0.312 
&   2.031 $ \pm$   0.399 \\  
     79 &   2.164 $ \pm$   0.368 &   0.072 $ \pm$   0.007 &   0.916 $ \pm$   0.363 &   2.585 $ \pm$   0.380 
&   5.069 $ \pm$   0.475 \\  
     96 &   2.487 $ \pm$   0.312 &  --0.005 $ \pm$   0.006 &   1.128 $ \pm$   0.314 &   3.116 $ \pm$   0.319 
&   1.920 $ \pm$   0.420 \\  
    149 &   1.705 $ \pm$   0.432 &   0.043 $ \pm$   0.007 &   1.164 $ \pm$   0.404 &   1.955 $ \pm$   0.430 
&   0.134 $ \pm$   0.556 \\  
\enddata
\end{deluxetable}
\clearpage

\clearpage

\clearpage


\clearpage


\clearpage

\clearpage

\clearpage

\clearpage

\clearpage

\clearpage

\clearpage
\clearpage

\clearpage

\clearpage


\clearpage


\end{document}